\documentclass[12pt]{article}
\usepackage{epsfig,amsfonts,amsbsy}
\newcommand{\affiliation}[1]{\textit{#1}}
\newtheorem{theorem}{Theorem}[section]

\makeatletter
\@addtoreset{equation}{section}
\makeatother

\newcommand{\ret}{\nonumber\\}

\newcommand{\rbk}[1]{\left(#1\right)}

\newcommand{\sumtwo}[2]%
{\mathop{\sum_{#1}}_{#2}}
\newcommand{\sumthree}[3]%
{\mathop{\mathop{\sum_{#1}}_{#2}}_{#3}}
\newcommand{\sumfour}[4]%
{\mathop{\mathop{\mathop{\sum_{#1}}_{#2}}_{#3}}_{#4}}
\newcommand{\suptwo}[2]%
{\mathop{\sup_{#1}}_{#2}}
\newcommand{\supthree}[3]%
{\mathop{\mathop{\sup_{#1}}_{#2}}_{#3}}
\newcommand{\supfour}[4]%
{\mathop{\mathop{\mathop{\sup_{#1}}_{#2}}_{#3}}_{#4}}
\newcommand{\inftwo}[2]%
{\mathop{\inf_{#1}}_{#2}}
\newcommand{\infthree}[3]%
{\mathop{\mathop{\inf_{#1}}_{#2}}_{#3}}
\newcommand{\inffour}[4]%
{\mathop{\mathop{\mathop{\inf_{#1}}_{#2}}_{#3}}_{#4}}

\newcommand{\calC}{{\cal C}}

\newcommand{\calE}{{\cal E}}

\newcommand{\calI}{{\cal I}}
\newcommand{\calJ}{{\cal J}}

\newcommand{\La}{\Lambda}
\newcommand{\up}{\uparrow}
\newcommand{\dn}{\downarrow}
\newcommand{\bs}{\backslash}

\newcommand{\cxs}{c_{x,\sigma}}

\newcommand{\Stot}{S_{\rm tot}}
\newcommand{\Sztot}{S_{\rm tot}^{(3)}}
\newcommand{\Smax}{S_{\rm max}}
\newcommand{\vac}{\Phi_{\rm vac}}

\newcommand{\Sopt}{\hat{\bf S}_{\rm tot}}

\renewcommand{\phi}{\varphi}

\newcommand{\Eup}{E_{\up}}
\newcommand{\Edn}{E_{\dn}}

\newcommand{\vsigma}{\boldsymbol{\sigma}}

\begin{document}

\begin{center}
{\bf An extension of Tasaki's flat-band Hubbard model }
\bigskip\\
Teppei Sekizawa

\affiliation{%
Department of  Physics, Tokyo Institute of Technology, 
Tokyo 152-8551, Japan
}
\end{center}

\date{\today}

\begin{abstract}
\baselineskip=1.5\baselineskip
 We present a new class of flat-band Hubbard 
 models
 which have saturated ferromagnetic ground
 states at two distinct electron numbers for 
 different values of parameters. 
 The models are extensions of Tasaki's
 flat-band models.
\end{abstract}



\section{Introduction}
 It is widely believed
 that the spin-independent Coulomb 
 interaction and the Pauli exclusion
 principle can generate ferromagnetism
 in itinerant electron systems.  
 One of the motivations to study the Hubbard 
 model has been to establish and understand
 the generation of ferromagnetism in simplified
 situations taking account of these effects
 \cite{Tasaki98}. Mielke \cite{Mielke} 
 and Tasaki \cite{Tasaki92}, independently, 
 presented  Hubbard models which
  exhibit saturated ferromagnetism for certain
  electron numbers when the Coulomb interaction $U$
  is positive.
 These models have a 
  common feature that the single electron spectra
  contain dispersionless bands, and are called
  flat-band models. In \cite{Tasaki95} and \cite{Tasaki03}, 
  Tasaki also discovered Hubbard models
  exhibiting ferromagnetism which models are nonsingular in the
  sense that both the density of states and the Coulomb interaction
  are finite. Recently
  Tanaka and Ueda succeeded in proving the existence
  of saturated ferromagnetism in a Hubbard model obtained
  by adding extra hopping terms to Mielke's flat-band model
  on the kagom\'e lattice \cite{Tanaka03}.

  Although the flat band Hubbard models are singular
  and not physically realistic, their study can be a basis
  of more realistic results about ferromagnetism. 
  It is therefore important to find out which flat-band
  models exhibit ferromagnetism. Although an abstract
  criterion was presented by Mielke \cite{Mielke2},
  we still do not know precise class of models
  which satisfy the criterion.

\begin{figure}[htbp]
\begin{center}
\includegraphics{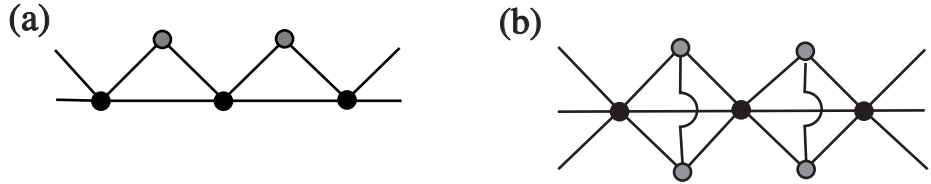}
\end{center}
\caption{The lattice structure and the hopping amplitude in the
one dimensional flat band models of (a)~Tasaki's and (b)~ours.
The black dots are the external sites (in $\cal E$) and
the gray dots are the internal sites (in $\cal I$).
Tasaki's model has one internal cite in each cell while
our has two.}
\label{onedim}
\end{figure}

  In this paper, we follow Tasaki's construction of his
  flat band models, and construct
   a new class of Hubbard models
  in arbitrary dimensions with finite $U$ and finite-range hopping.
   We prove that the models exhibit ferromagnetism in their
   ground states at two distinct electron numbers.
The difference between Tasaki's original model and ours 
can easily be seen from Fig.\ref{onedim}  where the simplest one-dimensional
versions of the models are illustrated. Tasaki's model has
one ``internal site'' (gray dot in the figure) in each unit cell,
while ours has two. This difference in lattice structure makes
our model to have different ``exchange mechanism'' where a single-electron
state localized at each pair of internal sites play an important role.
See section \ref{end} for more details.

We have thus found that a extension of Tasaki's construction
lead to a new class of models exhibiting ferromagnetism. 
We hope this study will shed light on general structure of
flat-band ferromagnetism.
\section{The model and main results}
\label{s:main}
\subsection{Construction of the lattice}
\label{s:lattice}
In the original flat-band models by Tasaki \cite{Tasaki92}, 
the basic cell in the lattice $\Lambda$ 
consists of a single internal site
and some external sites. 
In our new models,
the basic cell consists of two internal sites and some external sites. 
More precisely we let the basic cell be

\begin{equation}
C= \{u,v,x_1,x_2, \cdots ,x_n\}.
\label{eq:Cell}
\end{equation}
We call $u$ and $v$ the internal sites of $C$, 
and $x_1,x_2, \cdots, x_n$ the external sites. 

\begin{figure}[htbp]
\begin{center}
\includegraphics{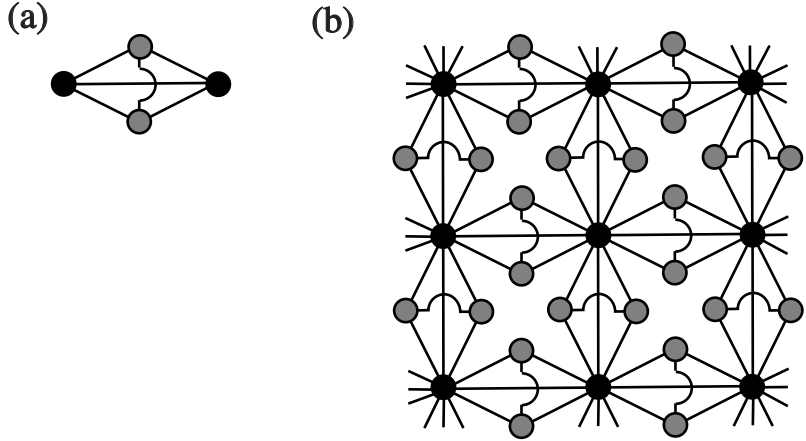}
\end{center}
\caption{An example of a cell and a lattice. From the quadrangular
cell with four sites (a), one can form (b) a decorated square lattice
by identifying four external sites.
This defines flat-band Hubbard model in two dimensions.
For the precise definition see (\ref{eq:Hh}) and (\ref{eq:H hop})
(where one should set $n=2$, $m=4$).
When $N_{\rm e}=| \Lambda |/5$, the model exhibits
   saturated ferromagnetism for any $t>0$, $s'>0$ and $U>0$.
   The model also exhibits
   saturated ferromagnetism for any $t>0$, $U>0$
   when $N_{\rm e}=3| \Lambda |/5$ and $s'=0$. See Theorem \ref{tfo:main}
and Theorem \ref{tft:main}. }
\label{cell}
\end{figure}

To form the lattice $\Lambda$, we assemble $M$ identical copies 
$C_1,C_2, \cdots ,C_{M}$ 
of the basic cell $C$, and identify external sites {from} $m$ distinct
cells and regard them as a single site.
In other words, an external site in $\Lambda$ is shared
by $m$ distinct cells. 
We denote by 
$| \Lambda |$ the number of sites in $ \Lambda $. 
See Fig.\ref{cell} for an example of a cell and a resulting lattice.

The lattice is naturally decomposed as
\begin{equation}
\Lambda=\cal I\cup \calE
\end{equation}
where $\cal I$ and $\cal E$ are the sets of internal sites
and external sites, respectively. We also denote by $\cal J$
 the assembly $\{1,2, \cdots, M\}$ of the indices of cells. 
{From} the above construction, we see
that the numbers of sites in these sublattices are
$|{\cal I}|=2M$, $|{\cal E}|=nM/m$.
By using $| \Lambda |$,
 we can write  $|{\cal I}|=2m| \Lambda |/(2m+n)$,
  $|{\cal E}|=n| \Lambda |/(2m+n)$ and $|{\cal J}|=m| \Lambda |/(2m+n)$.

In what follows we always regard $C_j$ as a subset of $\Lambda$. 
We denote the two internal sites in $C_j$ as $u_j$ and $v_j$. 
For an external site $x \in \calE$, we denote by $J_x$ 
the collection of indices $j$ such that $x \in C_j$. 
We also define $\La_x \subset \La$ to be the union of $m$ cells which contain
the site $x$.

\subsection{Fermion operators}
\label{s:fermi}
We consider an electron
 system on the lattice $\La$. 
For each site $r \in \La$ and $\sigma=\up,\dn$, 
we define the creation and the annihilation operators
$c^\dagger_{r,\sigma}$ and $c_{r,\sigma}$ for an electron at site
 $r$ with spin $\sigma$. These operators satisfy the canonical
 anticommutation relations
\begin{equation}
\{c^\dagger_{r,\sigma},c_{s,\tau}\}=\delta_{r,s}\delta_{\sigma,\tau},
\label{eq:crcs}
\end{equation}
and
\begin{equation}
\{c^\dagger_{r,\sigma},c^\dagger_{s,\tau}\}=\{c_{r,\sigma},c_{s,\tau}\}=
0,
\label{eq:crcscrcs}
\end{equation}
for any $r,s\in \La$ and $\sigma,\tau=\up,\dn$, where $\{A,B\}=AB+BA$. 
We denote by $\vac$ a normalized vector state which satisfies
 $c_{r,\sigma}\vac=0$ for any $r \in \La$ and $\sigma=\up,\dn$.
Then for arbitrary subsets $\La_{\up},\La_{\dn}\subset \La$, 
we define a state 
\begin{equation}
\rbk{\prod_{r\in \La_{\up}}c^\dagger_{r,\up}}
        \rbk{\prod_{r\in \La_{\dn}}c^\dagger_{r,\dn}}
        \vac \label{eq:psei}
\end{equation}
in which sites in $\La_{\up}$ are occupied by up-spin electrons
 and sites in $\La_{\dn}$ by down-spin electrons.

 Next we define total spin operator $\Sopt=(\hat S^{(1)}_{\rm tot},
\hat S^{(2)}_{\rm tot},\hat S^{(3)}_{\rm tot})$ by
\begin{equation}
\hat S^{(\alpha)}_{\rm tot}=\frac{1}{2}\sumtwo{r\in \La}
{\sigma,\tau=\up,\dn}c^\dagger_{r,\sigma}(p^{(\alpha)})_{\sigma,\tau}
c_{r,\tau}\label{ssss}
\end{equation}
for $\alpha=1,2,$ and $3$. Here $p^{(\alpha)}$ are the Pauli matrices
 defined by
\begin{equation}
p^{(1)}=\left(\begin{array}{cc}0&1 \\
1&0 \end{array} \right), \hspace{+1cm}
p^{(2)}=\left(\begin{array}{cc}0&-i \\
i&0 \end{array} \right), \hspace{+1cm}
p^{(3)}=\left(\begin{array}{cc}1&0 \\
0&-1 \end{array} \right).
\label{ppp}
\end{equation}

We finally define special fermion operators as in \cite{Tasaki92}. Let $\nu >0$
be a constant. For $x \in{\cal E}$, let
\begin{equation}
a_{x,\sigma}=c_{x,\sigma}-\frac{\nu}{2} \sum_{j \in J_{x}}
(c_{u_j,\sigma}+c_{v_j,\sigma}),
\label{eq:axs}
\end{equation}
where the sum is over $m$ sites adjacent to $x$. 
For $j \in {\cal J}$, let 
\begin{equation}
b_{j,\sigma}=c_{u_j,\sigma}+c_{v_j,\sigma}
+\nu\sum_{x\in C_{j}\cap \calE}\cxs,\label{eq:buos}
\end{equation}
where the sum is over the $n$ external sites in the cell $C_{j}$, and
\begin{equation}
d_{j,\sigma}=c_{u_j,\sigma}-c_{v_j,\sigma}.
\label{eq:duts}
\end{equation}

{From} the anticommutation relations for the basic $c$ operators, 
one can easily verify that
\begin{equation}
\{a^\dagger_{x,\sigma},b_{j,\tau}\}=
\{a^\dagger_{x,\sigma},d_{j,\tau}\}=
\{b^\dagger_{j,\sigma},d_{k,\tau}\}=0
\label{eq:abd}
\end{equation}
for any ${x \in{\cal E}},j,k \in {\cal J}$,
 and $\sigma ,\tau =\up ,\dn$.

The anticommutation relations for the $a$
 operators are
\begin{equation}
        \{a^\dagger_{x,\sigma},a_{y,\tau}\}=
        \cases{
        1+\displaystyle\frac{m\nu^2}{2},&if \( x=y \), \( \sigma=\tau \);\cr
\displaystyle\frac{\ell_{x,y}\nu^2}{4},&if \( x\ne y \), \( \sigma=\tau
\);\cr
0,&  otherwise.
        }\label{eq:aa}
\end{equation}
For $x,y \in\cal E$, we defined
\begin{equation}
\ell_{x,y}=|\La_{x}\cap\La_{y}\cap\calI|,
\label{eq:lxy}
\end{equation} 
which is the number of the internal sites directly
connected both to $x$ and $y$. 
For the \( b \) operators, we similarly have
\begin{equation}
\{b^\dagger_{j,\sigma},b_{k,\tau}\}=
\cases{
2+n\nu^2,&if \( j=k \), \( \sigma=\tau \);\cr
\ell_{j,k}\nu^2,&if \( j\ne k \), \( \sigma=\tau \);\cr
0,& otherwise.
}
\label{eq:bb}
\end{equation}
For \( j,k \in {\cal J}\), we defined
\begin{equation}
\ell_{j,k}=|C_{j}\cap C_{k}\cap\calE|,
\label{eq:luv}
\end{equation}
which is 
the number of external sites which are included in 
 both $C_j$ and $C_k$.
For the \( d \) operators, we have
\begin{equation}
        \{d^\dagger_{j,\sigma},d_{k,\tau}\}=
        \cases{
        2,&if \( j=k \), \( \sigma=\tau \);\cr
        0,&otherwise.
        }\label{eq:dd}
        \end{equation}

Since the states $a^\dagger_{x,\sigma} \vac,
 b^\dagger_{j,\sigma} \vac$ and
$d^\dagger_{j,\sigma} \vac$
 are linearly independent and
the number of these states are
$2| \Lambda |$,
 An arbitrary
many-electron state of the system
can be represented as a linear combination of
the basis states
\begin{eqnarray}
\Psi_{0}(\Eup,\Edn,J_{\up},J_{\dn},J'_{\up},J'_{\dn})=
       \ret && \hspace{-5.0cm}
        \rbk{\prod_{x\in E_{\up}}a^\dagger_{x,\up}}
        \rbk{\prod_{x\in E_{\dn}}a^\dagger_{x,\dn}}
        \rbk{\prod_{j\in J_{\up}}b^\dagger_{j,\up}}
        \rbk{\prod_{j\in J_{\dn}}b^\dagger_{j,\dn}}
        \rbk{\prod_{j\in J'_{\up}}d^\dagger_{j,\up}}
        \rbk{\prod_{j\in J'_{\dn}}d^\dagger_{j,\dn}}
        \vac \label{eq:pseipsei}
\end{eqnarray}
with arbitrary subsets
\( E_{\up},E_{\dn}\subset\calE \), 
\( J_{\up},J_{\dn},J'_{\up},J'_{\dn}\subset {\cal J}\). 
Here
\( |E_{\up}|+|E_{\dn}|+|J_{\up}|+|J_{\dn}|+|J'_{\up}|+|J'_{\up}|=N_{\rm e}
\)
 is the total electron number.

\subsection{An extension of Tasaki's flat-band model}
\label{s:exfl}
We study a Hubbard model with the Hamiltonian
\begin{equation}
H=t \sum\limits_{j=1}^{M} \sum_{\sigma=\up,\dn} b^\dagger_{j,\sigma}
b_{j,\sigma} +s' \sum\limits_{j=1}^{M} \sum_{\sigma=\up,\dn}
d^\dagger_{j,\sigma} d_{j,\sigma} +U \sum_{r \in \Lambda}
n^\dagger_{r,\up} n_{r,\dn}
\label{eq:H}
\end{equation}
where $t>0$, $s'$ and $U\ge 0$ are real , and
$n_{r,\sigma}=c^\dagger_{r,\sigma} c_{r,\sigma}$ is
the number operator.
We can rewrite the same Hamiltonian in the more standard form as
\begin{equation}
H=  \sumtwo{r,s \in \Lambda}{\sigma=\up,\dn} t_{r,s} 
c^\dagger_{r,\sigma}
c_{s,\sigma} +U \sum_{r \in \Lambda}
n^\dagger_{r,\up} n_{r,\dn}
\label{eq:Hh}
\end{equation}
where $t_{r,s}$ are the hopping amplitudes given by
\begin{equation}
        \cases{
        t_{x,x} = mt\nu^2,& \( x\in \calE \);\cr
        t_{u_i,u_i}=t_{v_i,v_i}=t+s',& \( i \in \calJ \);\cr
        t_{u_i,v_i}=t_{v_i,u_i}=t-s',& \( i \in \calJ \);\cr
        t_{x,u_i}=t_{u_i,x}=t_{x,v_i}=t_{v_i,x}= \cases{t\nu ,
        & \( x \in {\calC}_{i} \)
        \cr 0,& \( x \notin {\calC}_{i}\)}
        & \( x\in \calE \ , i \in \calJ  \);\cr
        t_{x,y}=\ell_{x, y} t\nu^2,& \( x,y \in \calE,x\neq y \);\cr
        t_{u_i,u_j}=t_{v_i,v_j}=t_{u_i,v_j}=0,& 
        \( i,j \in \calJ , i \neq j \).
        }\label{eq:H hop}
\end{equation}

{From} the anticommutation relations (\ref{eq:abd}) and (\ref{eq:dd}),
 one easily find that
the single-electron Schr\"odinger equation corresponding to (\ref{eq:H})
has $|{\cal E}|$-fold degenerate 
eigenstates $a_{x,\sigma}^\dagger\Phi_{\rm vac}$
 with energy $0$  and 
$M$-fold degenerate eigenstates $d_{j,\sigma}^\dagger\Phi_{\rm vac}$
with energy $2s^\prime$. 
It is also easy to see that the remaining eigenvalues are positive.
We have thus defined a new class of flat-band Hubbard models. 
The models exhibit ferromagnetism as the following two 
theorems state.

\begin{theorem}
\label{tfo:main}
Consider the above Hubbard model with
$N_{\rm e}=|{\cal E}|=n| \Lambda |/(2m+n)$ and $s'>0$. 
For any $U>0$, the ground states have total
spin $ \Stot= \Smax (=N_{\rm e}/2)$, 
and are non-degenerate apart {from} the
trivial $(2 \Smax +1)$-fold degeneracy. 
\end{theorem}

\begin{theorem}
\label{tft:main}
Consider the above Hubbard model with
$N_{\rm e}=|{\cal E}|+|{\cal J}|=(n+m)| \Lambda |/(2m+n)$ and $s'=0$. 
For any $U>0$, the ground states have total
spin $ \Stot= \Smax (=N_{\rm e}/2)$, 
and are non-degenerate apart {from} the
trivial $(2 \Smax +1)$-fold degeneracy.
\end{theorem}

It is remarkable that the new models show
saturated ferromagnetism
at two distinct electron numbers for different 
values of the parameters. This is
a unique property of our models.

\section{Proof}
\label{s:p}
We define the states $\Phi_{1\up},\Phi_{2\up}$ as
\begin{equation}
\Phi_{1\up}=\rbk{\prod_{x\in {\cal E}}a^\dagger_{x,\up}}\vac,\hspace{1cm}
\Phi_{2\up}=\rbk{\prod_{x\in {\cal E}}a^\dagger_{x,\up}}
  \rbk{\prod_{j\in {\cal J}}d^\dagger_{j,\up}} \vac.\label{phot}
\end{equation}
We decompose the Hamiltonian as $H=H_{\rm hop}+H_{\rm int}$ where
\begin{equation}
H_{\rm hop}=t \sum\limits_{j=1}^{M} \sum_{\sigma=\up,\dn} b^\dagger_{j,\sigma}
b_{j,\sigma} +s' \sum\limits_{j=1}^{M} \sum_{\sigma=\up,\dn}
d^\dagger_{j,\sigma} d_{j,\sigma}  ,
\label{eq:hhop}
\end{equation}

\begin{equation}
H_{\rm int}=U \sum_{r \in \Lambda} n^\dagger_{r,\up} n_{r,\dn}.
\label{eq:hint}
\end{equation}
Note that both $H_{\rm hop}$ and $H_{\rm int}$ are 
positive semidefinite.

\subsection{Proof of Theorem \ref{tfo:main}}
\label{s:po}
We consider  the case with
$N_{\rm e}=|{\cal E}|$ and $s'>0$,
 and prove Theorem $\ref{tfo:main}$. 
Since the proof is essentially the same
 as that found in \cite{Tasaki98, Tasaki92}, we shall be brief. 
Since $H\ge 0$ and $H \Psi_{1\up}=0$, we see
 that an arbitrary ground state $\Phi_{\rm GS}$ satisfies
\begin{equation}
H_{\rm hop}\Phi_{\rm GS}=0,\hspace{1cm} H_{\rm int}\Phi_{\rm GS}=0. 
\label{eq:phgso}
\end{equation}
{From} the second relation in (\ref{eq:phgso}), 
we further find that $\Phi_{\rm GS}$ must satisfy
\begin{equation}
c_{r,\up} c_{r,\dn}\Phi_{\rm GS}=0 \label{eq:crucrdgs}
\end{equation}
for any $r\in \Lambda$.

By using (\ref{eq:pseipsei}) and the first condition in 
(\ref{eq:phgso}), $\Phi_{\rm GS}$ can be
represented as a linear combination of the basis states
\begin{equation}
        \Psi_{1}(\Eup,\Edn,J_{\up},J_{\dn})=
        \rbk{\prod_{x\in E_{\up}}a^\dagger_{x,\up}}
        \rbk{\prod_{x\in E_{\dn}}a^\dagger_{x,\dn}}
        \vac.\label{eq:phrr}
\end{equation}

By using the anticommutation relations 
$\{c_{x,\sigma},a_{y,\tau}^\dagger\}
=\delta_{\sigma,\tau}\delta_{x,y}$, 
we see that
\begin{equation}
a^\dagger_{x,\dn}a^\dagger_{x,\up}
c_{x,\up}c_{x,\dn}
\Psi_{1}^{(\nu)}(\Eup,\Edn,J_{\up},J_{\dn})
=
\cases{
\Psi_{1}^{(\nu)}(\Eup,\Edn,J_{\up},J_{\dn}),
&if \( x\in\Eup\cap\Edn \);\cr
0,&otherwise,
}
\label{eq:aaccP2}
\end{equation}
for any $x \in \calE$. 
By using $(\ref{eq:crucrdgs})$ for $r \in \calE$ we find that 
only the basis states
 satisfying $\Eup\cap\Edn=\emptyset$, 
 contribute to $\Phi_{\rm GS}$.

In this way, $\Phi_{\rm GS}$ can be written as
\begin{equation}
\Phi_{\rm GS}=\sum_{\vsigma}
        g[\vsigma]
        \rbk{\prod_{x\in \calE}a^\dagger_{x, \vsigma (x)}}
        \vac \label{phgst}
\end{equation}
where the sum is over all the spin configuration $\vsigma=(\sigma_x)_{x \in
{\cal E}}$ on ${\cal E}$ and $g[ \vsigma]$ is
a coefficient.

By using (\ref{phgst}) and the anticommutation relations
$\{c_{u_{j},{\sigma}}, {a^\dagger_{x,{\tau}}}\}=-({\nu}/2){\delta
_{\sigma,\tau}} {\chi} {[x\in {\cal E}\cap{\calC_{j}}]}$, for
any $x \in {\cal E}$, where ${\chi} [\rm true]=1,{\chi} [\rm false]=0$,
 we get
\begin{eqnarray}
c_{u_{j},\up} c_{u_{j},\dn} \Phi_{\rm GS}=
\displaystyle\frac{{\nu}^2}{4}
\sumtwo {\alpha,\beta \in {\cal E}\cap \calC_{j}}
{{\rm s.t.}\alpha>\beta}
\sumtwo {\vsigma}{{\rm s.t.} \vsigma(\alpha)=\up,
 \vsigma(\beta)=\dn}
{\rm sgn}[\alpha,\beta]
(g[\vsigma]-g[{\vsigma}_{\alpha \leftrightarrow \beta}])
        \rbk{\prod_{x\in \calE \bs \{\alpha , \beta\}}a^\dagger_{x, \vsigma (x)}}
        \vac \label{eq:stst}
\end{eqnarray}
where we have introduced an arbitrary ordering in ${\cal E}$
to avoid double counting and
the factor ${\rm sgn}[\alpha,\beta]$ comes {from}
the exchange of fermion operators. The spin configuration
 ${\vsigma}_{\alpha \leftrightarrow \beta}$ is obtained 
from $\vsigma=(\sigma_\alpha)_{\alpha \in {\cal E}}$ by switching 
$\sigma_\alpha$ and $\sigma_\beta$. 
Since the basis states in $(\ref{eq:stst})$
 are all linearly independent, we find from the property $(\ref{eq:crucrdgs})$
that

\begin{equation}
g[\vsigma]=g[\vsigma_{\alpha \leftrightarrow \beta}]\label{eq:gg}
\end{equation}
for the sites $\alpha,\beta$
which belong to ${\cal E}\cap \calC_{j}$. 
Since the entire lattice
is connected, (\ref{eq:gg}) ensures that
the lowest energy is unique in
each sector with a fixed $\Sztot$. 
Therefore $\Phi_{1\up}$ is the
unique ground state apart {from} the
degeneracy for rotational invariance. 
This completes the proof of
Theorem \ref{tfo:main}.

\subsection{Proof of Theorem \ref{tft:main}}
\label{s:pt}
We treat the case with 
$N_{\rm e}=|{\cal E}|+|{\cal J}|$ and $s'=0$, 
and prove Theorem $\ref{tft:main}$. 
By using  $\Phi_{2\up}$ instead of $\Phi_{1\up}$, 
we find that the
conditions $(\ref{eq:phgso})$ and $(\ref{eq:crucrdgs})$ are still valid. 
By using (\ref{eq:pseipsei}) and the first condition in 
(\ref{eq:phgso}), 
we find that an arbitrary ground state $\Phi_{\rm GS}$ can be
represented as a linear combination of the basis states
\begin{equation}
\Psi_{2}(\Eup,\Edn,J_{\up},J_{\dn})=
        \rbk{\prod_{x\in E_{\up}}a^\dagger_{x,\up}}
        \rbk{\prod_{x\in E_{\dn}}a^\dagger_{x,\dn}}
        \rbk{\prod_{j\in J_{\up}}d^\dagger_{j,\up}}
        \rbk{\prod_{j\in J_{\dn}}d^\dagger_{j,\dn}}
        \vac.\label{eq:phrrrr}
\end{equation}
As before, $(\ref{eq:crucrdgs})$ and $(\ref{eq:aaccP2})$ imply
 that only the basis states with $E_{\up}\cap E_{\dn}=\emptyset$ 
 contribute to $\Phi_{\rm GS}$. 
 We will also prove that the basis states
 should satisfy $J_{\up}\cap J_{\dn}=\emptyset$ 
 in order to contribute to $\Phi_{\rm GS}$. In other words, $d$
 states cannot be doubly occupied in a ground state. 
We prove Theorem $\ref{tft:main}$
 assuming this claim. 

{From} (\ref{eq:phrrrr}) and
 the above mentioned constraints,
 we have
\begin{equation}
        \Phi_{\rm GS}=\sum_{\vsigma}
        g[\vsigma]
        \rbk{\prod_{x\in \calE}a^\dagger_{x, \vsigma (x)}}
        \rbk{\prod_{j\in \calJ}d^\dagger_{j, \vsigma (u_{j})}}
        \vac \label{phgstt}
\end{equation}
where the sum is over all the spin configurations $\vsigma=(\sigma_r)_{r \in
{\cal E}\cup \calJ}$  and $g[ \vsigma]$ is
a coefficient.Here, for notational simplicity, we identified the index set 
${\cal J}$ with the set of sites  $\{u_j\}_{j\in{\cal J}}$.  

By using (\ref{phgstt}) and the anticommutation relations
$\{c_{u_{j},{\sigma}}, {a^\dagger_{x,{\tau}}}\}=-({\nu}/2){\delta
_{\sigma,\tau}} {\chi} {[x\in {\cal E}\cap{C_{j}}]}$, 
$\{c_{u_{j},{\sigma}}, {d^\dagger_{k,{\tau}}}\}={\delta
_{\sigma,\tau}} {\chi} {[u_{k} \in {C_{j}}]}$ 
for any $x \in {\cal E}$, $j,k \in {\cal J}$, we get
\begin{eqnarray}
        c_{u_j,\up}c_{u_j,\dn}\Phi_{\rm GS}=
        \frac{\nu^2}{4}
        \sumtwo {\alpha , \beta\in{\cal E}\cap C_{j}}
        {s.t.\alpha>\beta}
        \sumthree {\vsigma}
        {s.t.{\vsigma_\alpha=\up}}{{ \vsigma_\beta=\dn}}
        {\rm sgn}(\alpha ,  \beta)
        (g[\vsigma]-g[\vsigma_{\alpha \leftrightarrow \beta}])
        \ret && \hspace{-5cm}
        \times
        \rbk{\prod_{x\in \cal E \bs \{\alpha , \beta\}}
        a^\dagger_{x, \vsigma (x)}}
        \rbk{\prod_{j\in \calJ}d^\dagger_{j, \vsigma (u_j)}}
        \vac 
        \ret && \hspace{-12cm}
        -\frac{\nu}{2}
        \sumtwo {\alpha \in{\cal E}\cap C_{j}}{\beta=u_j}
        \sumthree {\vsigma}
         {s.t.{\vsigma_\alpha=\up}} {{ \vsigma_\beta=\dn}}
        {\rm sgn}(\alpha ,  \beta)
        (g[\vsigma]-g[\vsigma_{\alpha \leftrightarrow \beta}])
        \ret && \hspace{-6cm}
        \times
        \rbk{\prod_{x\in \calE \bs \{\alpha\}}
        a^\dagger_{x, \vsigma (x)}}
        \rbk{\prod_{k\in \calJ \bs \{j\}}
        d^\dagger_{k, \vsigma (u_k)}}
        \vac.\label{eq:ccphss}
\end{eqnarray}

Since this quantity vanishes for all $j\in {\cal J}$, 
we finally find that
\begin{equation}
g[\vsigma]=g[\vsigma_{\alpha \leftrightarrow \beta}]. 
\label{eq:ggt}
\end{equation}
Since the entire lattice
is connected, (\ref{eq:ggt}) ensures that
the lowest energy state is unique in
each sector with a fixed $\Sztot$.
Therefore $\Phi_{2\up}$ is the
unique ground state apart {from} the
degeneracy for rotational invariance.
This completes the proof of
Theorem $\ref{tft:main}$.

It remains to prove  that $d$ states cannot
be doubly occupied. 
Since  $a$ states cannot be doubly occupied, 
the ground state can be expanded in the basis states $(\ref{eq:phrrrr})$ as  
\begin{equation}
        \Phi_{\rm GS}=\sumtwo {\Eup,\Edn,J_{\up},J_{\dn}}
        {\Eup\cap\Edn=\emptyset}
        f(\Eup,\Edn,J_{\up},J_{\dn})
        \Psi_{2}(\Eup,\Edn,J_{\up},J_{\dn}).
        \label{eq:psss}
\end{equation}

Let $u_{j}$ and $v_{j}$ be the internal sites
which belong to a cell $C_{j}$. 
By using (\ref{eq:psss}) and the anticommutation relations
$\{c_{u_{j},{\sigma}}, {a^\dagger_{x,{\tau}}}\}=-({\nu}/2){\delta
_{\sigma,\tau}} {\chi} {[x\in {\cal E}\cap{C_{j}}]}$, 
$\{c_{u_{j},{\sigma}}, {d^\dagger_{k,{\tau}}}\}={\delta
_{\sigma,\tau}} {\chi} {[u_k \in {C_{j}}]}$, 
$\{c_{v_{j},{\sigma}}, {d^\dagger_{k,{\tau}}}\}=-{\delta
_{\sigma,\tau}} {\chi} {[u_k \in {C_{j}}]}$ 
for any $x \in {\cal E}$, $j, k \in {\cal J}$, we get
\begin{eqnarray}
        c_{\gamma_{j},\up}c_{\gamma_{j},\dn}\Phi_{\rm GS}=
        \displaystyle\frac{{\nu}^2}{4}
        \sum_{x,y\in C_{j}}
        \sumthree {\Eup,\Edn,J_{\up},J_{\dn}}
        {s.t. x \in \Eup , y \in \Edn}
        {\Eup\cap\Edn=\emptyset}
        f(\Eup,\Edn,J_{\up},J_{\dn})
        {\rm sgn}(x,y)
        \Psi_{2}(\Eup\bs\{x\},\Edn\bs\{y\},J_{\up},J_{\dn})
        \ret && \hspace{-14.5cm}
        \mp \displaystyle\frac{{\nu}}{2}
        \sum_{x\in C_{j}}
        \sumthree {\Eup,\Edn,J_{\up},J_{\dn}}
        {s.t. x \in \Eup}
        {\Eup\cap\Edn=\emptyset}
        f(\Eup,\Edn,J_{\up},J_{\dn})
        {\rm sgn}(x,u_{j})
        \Psi_{2}(\Eup\bs\{x\},\Edn,J_{\up},J_{\dn}\bs\{j\})
        \ret && \hspace{-14.5cm}
        \mp \displaystyle\frac{{\nu}}{2}
        \sum_{y\in C_{j}}
        \sumthree {\Eup,\Edn,J_{\up},J_{\dn}}
        {s.t. y \in \Edn}
        {\Eup\cap\Edn=\emptyset}
        f(\Eup,\Edn,J_{\up},J_{\dn})
        {\rm sgn}(u_{j},y)
        \Psi_{2}(\Eup,\Edn\bs\{y\},J_{\up}\bs\{j\},J_{\dn})
        \ret && \hspace{-14.5cm}
        +
        \sumthree {\Eup,\Edn,J_{\up},J_{\dn}}
        {s.t. \Eup\cap\Edn=\emptyset}
        {j \in J_{\up}, j \in J_{\dn}}
        f(\Eup,\Edn,J_{\up},J_{\dn})
        {\rm sgn}(u_{j},u_{j})
        \Psi_{2}(\Eup,\Edn,J_{\up}\bs\{j\},J_{\dn}\bs\{j\})
         \label{eq:ccssss}
\end{eqnarray}
where $\mp$ is $-$ for $\gamma=u$ and $+$ for $\gamma=v$, and 
 ${\rm sgn}(x,y)$ depend on the sites $x,y$ and
 $\Eup,\Edn,J_{\up},J_{\dn}$.

These quantities vanish for all $j \in {\cal J}$ because of 
(\ref{eq:crucrdgs}). By adding these two relations thus obtained, 
we find that

\begin{eqnarray}
\sumthree {\Eup,\Edn,J_{\up},J_{\dn}}
        {s.t. \Eup\cap\Edn=\emptyset}
        {j \in J_{\up}, j \in J_{\dn}}
        f(\Eup,\Edn,J_{\up},J_{\dn})
        {\rm sgn}(u_{j},u_{j})
        \Psi_{2}(\Eup,\Edn,J_{\up}\bs\{j\},J_{\dn}\bs\{j\})
\ret && \hspace{-12.5cm}
+\displaystyle\frac{{\nu}^2}{4}
        \sum_{x,y\in C_{j}}
        \sumthree {\Eup,\Edn,J_{\up},J_{\dn}}
        {s.t. x \in \Eup , y \in \Edn}
        {\Eup\cap\Edn=\emptyset}
        f(\Eup,\Edn,J_{\up},J_{\dn})
        {\rm sgn}(x,y)
        \Psi_{2}(\Eup\bs\{x\},\Edn\bs\{y\},J_{\up},J_{\dn})=0
\label{eq:stdfst}
\end{eqnarray}

The linear independence of $\Psi_{2}$ implies that
each coefficient of $\Psi_{2}$ in (\ref{eq:stdfst}) is vanishing. 
This means that

\begin{eqnarray}
        f(\Eup,\Edn,J_{\up},J_{\dn})
        {\rm sgn}(u_{j},u_{j})=
\ret && \hspace{-3cm}
-\displaystyle\frac{{\nu}^2}{4}
        \sum_{x,y\in C_{j}}
        \!\!
        \sumfour {\Eup',\Edn',J_{\up}',J_{\dn}'}
        {\Eup'\cap\Edn'=\emptyset}
        {s.t.\{\Eup'\bs\{x\},\Edn'\bs\{y\},J_{\up}',J_{\dn}'\}}
        {= \{\Eup,\Edn,J_{\up}\bs\{j\},J_{\dn}\bs\{j\}\}}
        \!\!
        f(\Eup',\Edn',J_{\up}',J_{\dn}')
        {\rm sgn}(x,y)
\label{eq:fsps}
\end{eqnarray}
for any $\Eup,\Edn\subset \cal E$ and $J_{\up},J_{\dn}\subset
 {\cal J}$ with $\Eup\cap\Edn=\emptyset$. 
Note that the two electrons occupying the same $d$ state in
 $\{\Eup,\Edn,J_{\up},J_{\dn}\}$ are removed and added
  to $\cal E$ in the new configuration $\{\Eup',\Edn',J_{\up}',J_{\dn}'\}$. 

Now suppose that $d$ state is doubly occupied in $r$ different 
cells. More precisely we assume that there is a configuration 
$\{\Eup,\Edn,J_{\up},J_{\dn}\}$ with $f(\Eup,\Edn,J_{\up},J_{\dn})\neq 0$
 such that $|J_{\up}\cap J_{\dn}|=r$. We use the relation
 (\ref{eq:fsps}) to this configuration by setting $j \in J_{\up}\cap J_{\dn}$.
  Then one finds that the summation in the right-hand side of (\ref{eq:fsps}) 
 is over configurations $\{\Eup',\Edn',J_{\up}',J_{\dn}'\}$
 where the $d$ state in $C_j$ is not occupied  and 
 there are two more occupied $a$ states 
  compared with the original configuration. 
 Thus $\{\Eup',\Edn',J_{\up}',J_{\dn}'\}$ has $(r-1)$ cells
 with doubly occupied $d$ states and at least 
 one cell with the empty $d$ state.
  Because of (\ref{eq:fsps}), there is
  at least one such configuration with non vanishing  
  $f(\Eup',\Edn',J_{\up}',J_{\dn}')$.
 Repeating the argument $r$ times, one concludes that 
 there is at least one configuration $\{\Eup',\Edn',J_{\up}',J_{\dn}'\}$
 with $f(\Eup',\Edn',J_{\up}',J_{\dn}')\neq 0$ 
 which has no cells with doubly occupied $d$ states 
 and at least $r$ cells with no $d$ states.  But
  this is a contradiction since the maximum possible
  electron number for such a configuration is 
  $|{\calE}|+|{\cal J}|-r<|{\calE}|+|{\cal J}|=N_{\rm e}$. 
This prove the desired claim that $r$ is always $0$.

\section{Discussions}
\label{end}
Let us make two remarks about our model.

\begin{figure}[htbp]
\begin{center}
\includegraphics{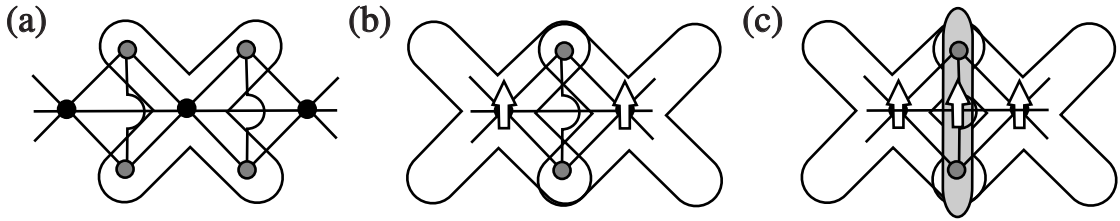}
\end{center}
\caption{Exchange mechanisms in the one dimensional models.
(a)~An almost localized state for electrons in the lowest flat-band ($a$
state).
(b)~In the case of Theorem \ref{tfo:main}, overlap of two states almost
localized at neighboring external sites generate ferromagnetism
just as in Tasaki's flat-band model.
(c) In the case of Theorem \ref{tft:main},
the exchange interaction also involves a state localized at two 
internal sites.}
\label{ferro}
\end{figure}

First we discuss the mechanisms that generate
ferromagnetism in the present and Tasaki's models.
In Tasaki's models, electrons in the lowest flat-band
may be regarded (in the basis corresponding to the $a$ operators)
as almost localized at external sites. Roughly speaking, a small overlap of  
the wave functions at an internal site generates  
``exchange interaction'' which leads to ferromagnetism 
as in Fig.\ref{ferro} (a). In the situation of Theorem \ref{tfo:main},
 the picture is almost the same in our models. The electrons  
are almost localized at external sites and overlap at intermediate
sites as in Fig.\ref{ferro} (b). In the situation of Theorem \ref{tft:main},
however, the picture is essentially different from that in Tasaki's model.
Each electron in the lowest flat bands is either almost localized
at an external site or localized at a pair of internal sites. The basic
``exchange interaction'' involves 
three electrons as in the Fig.\ref{ferro} (c).
This is why the proof of Theorem \ref{tft:main} required a new technique.

Secondly let us discuss the possibility of further extending Tasaki's
construction. A natural question is whether one can treat models 
with three or more internal sites. As for results corresponding to 
Theorem \ref{tfo:main}, it is obvious that our proof (and Tasaki's
original proof) automatically extends to such models.
But results corresponding to Theorem \ref{tft:main}, which involves
a new exchange mechanism, are much more delicate. We suspect
that a new idea is required to cover the general cases.

\bigskip 

Part of the present work was done when the author was at Department of  
Physics, Gakushuin University.
I wish to thank Hal Tasaki for introducing me to the problem  
of ferromagnetism and for various useful discussions and comments,  
Akinori Tanaka for indispensable comments, suggestions and  
discussions.  I also thank Masahito Ueda for useful discussions  
and continual encouragements.



\begin{thebibliography}{99}
\bibitem{Tasaki98}
H. Tasaki, Prog. Theor. Phys. \textbf{99}, 489 (1998) and references therein.
\bibitem{Mielke}
A. Mielke, J. Phys. A: Math. Gen. \textbf{24}, L73 (1991); \textbf{24}, 3311(1991);
\textbf{25}, 4335 (1992).
\bibitem{Tasaki92}
H. Tasaki, Phys. Rev. Lett. \textbf{69}, 1608 (1992).
\bibitem{Tasaki95}
H. Tasaki, Phys. Rev. Lett. \textbf{75}, 4678 (1995).
\bibitem{Tasaki03}
H. Tasaki,  to appear in Comm. Math. Phys, cond-mat/0301071.
\bibitem{Tanaka03}
        A. Tanaka and H. Ueda,  Phys. Rev. Lett. \textbf{90}, 067204 (2003).
\bibitem{Mielke2}
A. Mielke, Phys. Lett. A: \textbf{174}, 443 (1992).


\end{thebibliography}
\end{document}